\documentclass[12pt]{article} 
\pdfoutput=1
\usepackage{epsfig,graphicx,}
\usepackage{amsmath,amsfonts,amssymb,stmaryrd,latexsym}
\usepackage{url,subcaption,comment}
\newcommand{\be}{\begin{equation}}
\newcommand{\ee}{\end{equation}}
\newcommand{\bea}{\begin{eqnarray}}
\newcommand{\eea}{\end{eqnarray}}
\newcommand{\bear}{\begin{eqnarray}}
\newcommand{\eear}{\end{eqnarray}}
\newcommand{\ba}{\begin{array}}
\newcommand{\ea}{\end{array}}

\newcommand{\eg}{{\it e.g.}}
\newcommand{\ie}{{\it i.e.}}

\usepackage[usenames,dvipsnames]{xcolor}

\usepackage{dcolumn}
\usepackage{slashed}

\newcommand{\met}{\slashed{E}_T}

\newcommand{\lsim}{\!\mathrel{\hbox{\rlap{\lower.55ex \hbox{$\sim$}} \kern-.34em \raise.4ex \hbox{$<$}}}}
\newcommand{\gsim}{\!\mathrel{\hbox{\rlap{\lower.55ex \hbox{$\sim$}} \kern-.34em \raise.4ex \hbox{$>$}}}}

\newcommand{\vev}[1]{ \left\langle {#1} \right\rangle }
\newcommand{\abs}[1]{ \left| {#1} \right| }
\def\GeV{\text{ GeV}}
\def\TeV{\text{ TeV}}
\def\MeV{\text{ MeV}}
\def\fb{\text{ fb}}
\def\pb{\text{ pb}}

\newcommand{\Sref}[1]{Section~\ref{#1}}

\newcommand{\Fref}[1]{Figure~\ref{#1}}
\newcolumntype{M}{>{$}c<{$}} 

\begin{document}

\baselineskip=18pt \pagestyle{plain} \setcounter{page}{1}

\vspace*{-1cm}

\noindent \makebox[11.5cm][l]{\small \hspace*{-.2cm} }{\small Fermilab-Pub-16-190-T}  \\  [-1mm]

\begin{center}

{\large \bf   Multi-step production of a diphoton resonance 
} \\ [9mm]

{\normalsize \bf Bogdan A. Dobrescu, Patrick J. Fox and John Kearney \\ [3mm]
{\small {\it
Theoretical Physics Department, Fermilab, Batavia, IL 60510, USA   
  }}\\
}

\center{May 27, 2016; Revised November 17, 2016}

\end{center}

\vspace*{0.2cm}

\begin{abstract}
Among the questions that would be raised by the observation of a new resonance at the LHC, 
particularly pressing are those concerning the production mechanism:
What is the initial state?
Is the resonance produced independently or in association with other particles?
Here we present two weakly-coupled renormalizable models for production of a diphoton resonance that differ in both their initial and final states. In one model, a scalar particle produced through gluon fusion decays into a diphoton particle and a light, long-lived pseudoscalar.
In another model, a $Z'$ boson produced from the annihilation of a strange-antistrange quark pair undergoes a cascade decay that leads to a diphoton particle and two sterile neutrinos.
Various kinematic distributions may differentiate these models from the canonical model where a diphoton particle is directly produced in gluon fusion.
\end{abstract}


{\small
\tableofcontents
}

\section{Introduction} \setcounter{equation}{0}

In Run 2 of the LHC, started in 2015,  the proton-proton collisions at a center-of mass energy of 13 TeV 
open a large new window towards the laws of nature at the shortest distances directly accessible so far.
In particular, the existence of resonances indicative of physics beyond the Standard Model (BSM) will be tested 
up to higher masses and smaller cross sections than ever before.

Upon the observation of such a resonance, an important question will be how the new particle was produced. More precisely, it would be crucial to find out what is the initial state that produced the resonance, and whether 
the resonance is produced by itself or in association with other particles.
Features of the signal events, including the amount of missing transverse energy or unusual jet activity, may indicate the presence of additional states. 
However, it is possible that the presence of some final state particles  is obscured by small mass splittings within cascade decays, leading to a signal being na\"ively interpreted at first as a singly-produced resonance.

The excess $\gamma\gamma$ events observed by the ATLAS and CMS experiments using the 2015 data \cite{ATLAS-CONF-2015-081,CMS:2015dxe,ATLAS-CONF-2016-018,CMS:2016owr}, which were subsequently 
attributed to large statistical fluctuations \cite{ATLAS:2016eeo, Khachatryan:2016yec}, have highlighted the importance of these questions about production in the context of a diphoton resonance.
Many theoretical studies have been devoted to a ``canonical'' interpretation of a diphoton peak, in which a scalar particle produced from a gluon-gluon initial state that decays  directly into two photons (for reviews, see \cite{Staub:2016dxq,Franceschini:2016gxv}, for earlier work see, \eg, \cite{Fox:2011qc}).
However, the possibility that a high-mass diphoton resonance be produced in association with other particles, particularly ones that are difficult to observe, is relatively unexplored.

In this paper we discuss alternative models that lead to $\gamma\gamma$ resonances, in which additional final state particles generate only a small amount of missing energy, such that the signal origin could be misinterpreted as the canonical model. We highlight kinematic distributions that may be useful in distinguishing between the models, and discuss model-building challenges.

Specifically, we present a couple of renormalizable models in which a diphoton resonance arises from cascade decays of some slightly heavier particles.
In the first model, a scalar particle produced in gluon fusion decays into two pseudoscalars, 
one of which ($A$) is the diphoton resonance and the other ($A'$) is very light and quasi-stable  (we refer to this as a 2-step model). 
In the second model, a $Z'$ boson couples to the right-handed $s$ and $b$ quarks as well as to two new fermions, $N$ and $\nu_s$, 
which are singlets under the SM gauge group. One of the  $Z'$  decay modes is into $\bar \nu_s N$, with a subsequent decay of $N$ into $\nu_s$ and the diphoton resonance (we refer to this as a 3-step model). 
We explore how the kinematic distributions could eventually differentiate between these models.

Two-step production topologies\footnote{Our model has little in common with the ``2-step decay" studied in \cite{Cho:2015nxy}. 
Also, the fact that $A$ and $A'$ are different particles (with a large mass splitting) distinguishes our 2-step model from
the model where two diphoton particles are simultaneously produced \cite{Huang:2015evq,Altmannshofer:2015xfo}.} 
were previously considered in \cite{Franceschini:2015kwy, Bernon:2016dow}, but only in the context of simplified models and with 
focus on regions of parameter space in which the presence of additional final state particles would be immediately apparent.
A more complete 2-step model is discussed in \cite{Badziak:2016cfd} in the context of the NMSSM; again, the presence of a couple of $b$ jets in each diphoton event 
would clearly distinguish such a model from the canonical model.

Our renormalizable 2-step model naturally accommodates the small mass splitting between the  scalar ($\varphi$) produced in gluon fusion and the pseudoscalar $A$ that decays into photons by embedding them into the same complex scalar field.
The pseudoscalar $A'$ that escapes the detector is naturally very light because it is the pseudo-Nambu Goldstone boson associated with a $U(1)$ symmetry. 
Consequently, in spite of the presence of an additional particle in the final state, missing energy is suppressed, and it may at first appear that the resonance was produced by itself.

The 3-step model is more peculiar, because the $s$-channel resonance is a spin-1 particle 
(an alternative model is discussed in \cite{Altmannshofer:2015xfo}) and the initial state is mostly a strange-antistrange pair.
Even though some mass splittings are assumed to be small without a symmetry reason, this model
is interesting because the initial state does not involve gluon fusion, and so it leads to different pattens in QCD radiation and hence jet multiplicities. Meanwhile, sea quark production exhibits a $\sqrt{s}$ dependence
for the  production cross section similar to that for the gluon-initiated process, consistent with the potential misidentification of the resonance as arising from gluon fusion.

In \Sref{sec:2step} we present the 2-step scalar model and discuss its phenomenological implications. 
In \Sref{sec:3step} we construct and analyze the 3-step $Z'$ model. Examples of the kinematic 
distributions predicted in these models, as well as in the canonical ``1-step" model, are shown in \Sref{sec:distributions}. 
We also compare to the ATLAS kinematic distributions provided in \cite{ATLAS-CONF-2016-018} to demonstrate how even limited data could help distinguish 
the canonical model from the multi-step models. 
Our conclusions are presented in \Sref{sec:conclusions}.

\bigskip

%
\section{A 2-step scalar model} \setcounter{equation}{0}
\label{sec:2step}

In this section we present a simple model in which the spin-0 particle produced in gluon fusion is 
different than the spin-0 particle that decays into a photon pair. 
The model consists of two complex scalar fields, $\phi$ and $\phi'$, which are SM singlets, a  real scalar field $\Theta^a$ 
that is a color-octet ($a = 1, ...,8$), and a lepton $\psi$ of electric charge one, which is vectorlike with respect to the SM gauge group. 

The SM-singlet scalars are formed of the following real scalar fields:
\bear
&& \phi =  \langle \phi \rangle +  \frac{1}{\sqrt{2}} \left( \varphi + i A \right)  ~~,
\nonumber \\  [2mm]
&& \phi' =  \langle \phi' \rangle + \frac{1}{\sqrt{2}} \, \left( \varphi'  + i A^{\prime } \right)   ~~,
\eear
We consider the case where the VEVs satisfy 
$\langle \phi' \rangle \gg \langle \phi \rangle > 0$. We will neglect the $\phi$ VEV for now, and we will check later that this is a fair approximation.
The masses of the two CP-even scalars $\varphi$ and $\varphi'$, $M_\varphi$ and $M_{\varphi'}$, satisfy  $M_\varphi < M_{\varphi'}$; 
$\varphi'$ will not be important in what follows.
The masses of the two CP-odd scalars $A$ and $A^{\prime }$ satisfy  $M_{A'} \ll M_A$ and 
 $M_\varphi  > M_A + M_{A'}  $. The CP-odd scalar $A$ will play the role of the diphoton resonance at the TeV scale.

The interactions of these spin-0 particles will be selected such that the cascade decay 
$\varphi  \to A' A \to A' \gamma\gamma$ has a large branching fraction, and $A'$ does not decay inside the detector.

\subsection{Scalar interactions}
\label{subsec:scalarints}

We assume that  $\phi$ interacts with the color-octet scalar,  $\Theta^a$, via the following CP-conserving term  in the Lagrangian:
\be
- \frac{\kappa}{2} \, \phi' \phi \, \Theta^a \Theta^a + {\rm H.c.}   \supset    - \frac{\kappa \langle \phi' \rangle}{ \sqrt{2}}  \, \varphi \, \Theta^a \Theta^a    ~~,
\label{eq:kappa}
\ee
where $\kappa$ is a real\footnote{We impose CP symmetry on the term (\ref{eq:kappa})  in order to avoid an $A\Theta^a \Theta^a$ coupling that would 
produce an $s$-channel diphoton resonance. If $\kappa$ had a complex phase $\alpha_\kappa$, then the ratio of $A$ and $\varphi$ production cross sections would be $\tan^2\!\alpha_\kappa$. } dimensionless parameter.
This coupling leads at one loop to $gg \to \varphi$ production at the LHC \cite{Boughezal:2010ry,Dobrescu:2011aa}. 
Other terms in the potential that involve $\Theta^a$ are given by
\be
V(\Theta) = \frac{1}{2} \left( M_\theta^2 + \kappa_1 |\phi|^2 + \kappa_2  |\phi'|^2 \right) \Theta^a \Theta^a + \frac{\lambda_\Theta}{8} \left( \Theta^a \Theta^a \right)^2 + \mu_\Theta \, d_{abc} \Theta^a \Theta^b \Theta^c ~~.
\label{Vtheta}
\ee
Here $\lambda_\Theta >0$ is a dimensionless parameter relevant in \Sref{subsec:lhc2step}, and  $\mu_\Theta$
is a trilinear coupling that allows the decay $\Theta \to gg $ at one loop \cite{Bai:2010dj}. The first term in  $V(\Theta)$ leads to a squared-mass for the color-octet particle,
$M_\Theta^2 \simeq M_\theta^2 + \kappa_2  \langle \phi' \rangle^2$, which we take to be positive.

\begin{figure}[t] 
\vspace*{-2.5cm}
   \centering
    \includegraphics[width=1\textwidth, angle=0]{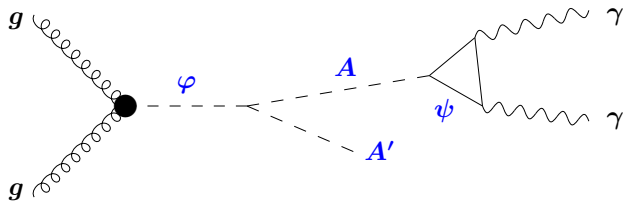}
\vspace*{-7.5cm}
   \caption{Diagram for $A$ production at the LHC in the 2-step model, where $A$ is a heavy CP-odd scalar decaying into two photons. 
   $A'$ is a light pseudoscalar that escapes the detector. The black disk represents the sum over loop contributions from the color-octet scalar $\Theta$.  }
   \label{fig:phiAdiagram}
\end{figure}

The $\phi$ and  $\phi'$ fields have a dimension-4 coupling in the potential,
\be
V_\lambda = \frac{\lambda}{4} \, (\phi \,  \phi' )^2  +   {\rm H.c.}   ~~,
\label{eq:quartic}
\ee
where again we impose CP symmetry so that $\lambda$ is a real dimensionless parameter (implying that the $A\to A'A'$ decay is negligible).
The above term includes the following interaction of $\varphi $ with the CP-odd scalars in the Lagrangian:
\be
\frac{\lambda \langle \phi' \rangle }{\sqrt{2}} \, \varphi  \, A A^{\prime }  ~~.
\ee
As a result, the  $\varphi $ scalar decays into $A A'$ with a width
\be
\Gamma (\varphi  \to A A^{\prime } ) = \frac{\lambda^2 \langle \phi' \rangle^2}{32 \pi M_\varphi}  
\left[ \left(1+ \frac{M_A} {M_\varphi} \right)^{\! 2} - \frac{M_{A'}^2} {M_\varphi^2} \right]^{\! 1/2} 
\left[ \left(1- \frac{M_A} {M_\varphi} \right)^{\! 2} - \frac{M_{A'}^2} {M_\varphi^2} \right]^{\! 1/2}  
~~.
\label{eq:phiAA}
\ee
Even though this width is phase-space suppressed, the $\varphi  \to A A^{\prime }$ branching fraction 
is large because the only other significant decay mode of $\varphi $, into two gluons, is loop-suppressed; a quantitative assessment is postponed until \Sref{subsec:lhc2step}. 
The primary ingredient necessary for associated production of a diphoton peak is thus in place:  the ``2-step production"  $gg \to \varphi \to A A'$. 
The subsequent $A\to \gamma \gamma$ decay, discussed in \Sref{subsec:Abrs}, then gives rise to a diphoton signal at the LHC (see the diagram in \Fref{fig:phiAdiagram}).

The scalar interactions introduced so far, and below, exhibit a (spontaneously-broken) global $U(1)$ symmetry under which $\phi$ and $\phi'$ rotate oppositely.  The full scalar potential allowed by this symmetry is
\be
V(\phi,\phi') = M_0^2 |\phi|^2 - M_0'^{2} |\phi'|^2 - (b^2 \phi \phi' + {\rm H.c.}) + \frac{\lambda_1}{2}  |\phi|^4 +  \frac{\lambda_2}{2} |\phi'|^4 + \lambda_3 |\phi \phi'|^2 + V_\lambda ~~.
\label{eq:Vtheta}
\ee
We assume that all parameters are real and positive and thus any VEVs are also positive.  In addition we take 
$b^2 \ll  M_0^2$ so that $\langle \phi' \rangle \approx M'_0 / \sqrt{\lambda_2} \gg \langle \phi \rangle$. 
Note that  the VEV of $\phi$ is induced by the $b^2$ term,
\be
\langle \phi \rangle \approx \frac{b^2 \langle \phi' \rangle}{M_0^2 + ( \lambda_3 + \lambda/2)  \langle \phi' \rangle^2} ~~.
\ee  
In the $b/M_0 \to 0$  limit, the Nambu-Goldstone boson arising from the spontaneously broken $U(1)$ is the $A'$ component of $\phi'$. That limit, however, is unstable because 
the coupling to $\Theta$ in Eq.~(\ref{eq:kappa}) induces a 1-loop contribution to $b^2$. For $b^2 \neq 0$, the Nambu-Goldstone boson is a linear combination of $A'$ and $A$.
In practice it is sufficient to have  $b/M_0 \lsim 0.3$, as the corrections to the $\varphi$ branching fractions and to the $M_\varphi - M_A$ mass splitting are only of order 
$b^4/M_0^4$. We will neglect these corrections, so we will keep the $A'$ notation for the Nambu-Goldstone boson.

With the Lagrangian introduced so far, $A'$ remains strictly massless.
A small $A'$  mass can be easily induced by including an explicit breaking of the global $U(1)$, for example a 
$M_{A'}^2 \phi^{\prime 2}$ term with $ M_{A'} \ll  |M'_0|$.
As discussed in \Sref{subsec:lhc2step}, values of $M_{A'}$ as low as 6 MeV (and perhaps lower) are phenomenologically viable.

The quartic interaction $V_\lambda$ also leads to a mass splitting of $\varphi$ and $A$:
\be
M_{\varphi,A}^2 = M_0^2 + \left( \lambda_3 \pm \frac{\lambda}{2} \right) \langle \phi' \rangle^2      ~~,
\ee
so that  the mass-squared difference is
\be
M_\varphi^2 - M_A^2 = \lambda \langle \phi' \rangle^2  ~~.
\label{eq:lambda}
\ee
The $V_\lambda$ term can be induced at 1-loop from the coupling to $\Theta$, thus in the absence of fine tuning one would expect that $ \lambda \gsim \kappa^2/(8\pi^2)$.
At the same time, in order to suppress missing energy in the final state, we require $M_\varphi-M_A\lsim O(50) \GeV$, so
that $\lambda$ must be smaller than about $0.12 M_\varphi^2/ \langle \phi' \rangle^2$.

\subsection{Branching fractions of $A$}
\label{subsec:Abrs}

To generate a diphoton signal, we introduce interactions allowing the decay $A \to \gamma \gamma$. This can be accomplished by coupling 
the scalars $\phi$ and $\phi'$  to the vectorlike lepton $\psi$, 
\be
- y_\psi \, \phi \, \bar \psi_L \psi_R   - y'_{\psi} \,  \phi' \, \bar \psi_R \psi_L +   {\rm H.c.}   ~~,
\label{eq:Yukawa2step}
\ee
where the Yukawa couplings are $y_\psi,  y'_{\psi}  >0$.
We assign global $U(1)$ charges to $\psi_L$ and $\psi_R$ such that the above terms are  $U(1)$ invariant.
As a result, a mass term for $\psi$ can be generated only by the $U(1)$ breaking VEVs. Thus, the vectorlike lepton has a mass
\be
m_\psi \simeq  y'_{\psi}  \langle \phi' \rangle  ~~.
\label{eq:mpsi}
\ee

The Yukawa interactions (\ref{eq:Yukawa2step}) contribute at one loop to the $b^2$ term in $V(\phi,\phi')$.
Hence, the values of the VEVs shift slightly. We neglect the effects of the $\psi$ loops, as they are not larger than those of the $\Theta$ loops discussed in \Sref{subsec:scalarints}.
As before, $A'$ remains a massless Nambu-Goldstone boson unless we choose to include explicit $U(1)$ breaking terms.

The first term in Eq.~(\ref{eq:Yukawa2step}) includes the following interaction:
\be
- y_\psi \, i A \bar \psi \gamma_5 \psi  ~~.
\ee
For $m_\psi > M_{A}/2$, the vectorlike lepton can be integrated out leading to a dimension-5
interaction of the CP-odd scalar $A$ with two 
 SM gauge bosons. If $\psi$ is an $SU(2)_W$ singlet of hypercharge $+1$, the  
dimension-5 interaction is given by
\be
\frac{\alpha \, y_\psi}{8\sqrt{2} \pi \, m_\psi  \cos^2\!\theta_w } \; A  \, B^{\mu\nu}  \widetilde B_{\mu\nu}
\label{eq:hyper}
\ee
where $\tilde{B}^{\mu \nu} = \frac{1}{2} \epsilon^{\mu \nu \alpha \beta} B_{\alpha \beta}$.
The resulting width for the $A$ decay into photons is given by
\be
\Gamma (A \to \gamma\gamma) = \frac{\alpha^2 \, y_\psi^2 \, M_A^3 }{128\pi^3 \, m_\psi^2}  ~~,
\label{eq:diphotonWidth}
\ee
with the electromagnetic coupling constant $\alpha$ evaluated at a scale of order $M_A$. 
Besides couplings to photons, the operator~(\ref{eq:hyper}) includes $A$ interactions with $Z\gamma$ and  $ZZ$, which give the following 
decay widths:
\bear
&&  \Gamma (A \to Z \gamma) =  2 \tan^2\!\theta_w  \; \Gamma (A \to \gamma\gamma)   ~~,
\nonumber \\  [2mm]
&&   \Gamma (A \to Z Z) =  \tan^4\!\theta_w \; \Gamma (A \to \gamma\gamma) ~~.
\eear
Here $\theta_w$ is the electroweak mixing angle at the $M_\varphi$ scale, so that $ \tan^2\!\theta_w \approx 0.30$.
These subdominant channels offer alternative methods for confirming the existence and nature of the resonance. Here, we have taken $\psi$ to be an $SU(2)_W$ singlet such that the decay to diphotons dominates---other choices would lead to different branching fractions and, for non-singlet representations, a decay to $W^+ W^-$ would also be relevant.

The quartic coupling of $\phi$  to $\Theta$ and $\phi'$ shown in Eq.~(\ref{eq:kappa}) includes a 
\be
- \frac{\kappa}{2} A A'  \, \Theta^a \Theta^a 
\label{eq:AAgg}
\ee
interaction. Consequently, the $A$ particle does not decay into gluons but it has a 1-loop, 3-body decay into  $A' gg $: 
\be
\Gamma (A \to A'gg) \simeq
 \frac{\alpha_s^2 \, \kappa^2  \, M_A^5 }{6 \, (8\pi)^5 M_\Theta^4 }   ~~.
 \ee
There is an additional contribution to the $A\to A' gg $ amplitude from an off-shell $\varphi$, which interferes with the one due to Eq.~(\ref{eq:AAgg}). However, this
is a subdominant contribution when $M_\varphi - M_A \ll M_A$, and it can be safely neglected here.
We expect that higher-order QCD corrections 
enhance $\Gamma (A \to\! A'gg)$ by a factor of order 2.

Let us compute the  widths for some benchmark points in the parameter space.
We fix the couplings 
\be
\lambda = 0.1 \;\; , \;\;      y_\psi = y'_\psi = \lambda_\Theta =  1   ~~,
\label{eq:coupling-values}
\ee
and then we define benchmark point 1 by
\bear
&&
 M_A = 750 \; {\rm GeV}   \;\; , \;\;    \langle \phi'\rangle = 680 \; {\rm GeV} \; \Rightarrow  \; M_\varphi = 780 \; {\rm GeV} ,  \;  m_\psi = 680   \; {\rm GeV}   ~~,
 \nonumber \\ [2mm]
&&   M_\Theta = 800   \; {\rm GeV}    \;\; , \;\;    \kappa =  0.5   ~~,
\label{eq:bench}
\eear
and benchmark point 2 by
\bear
&&
 M_A = 1.5 \; {\rm TeV}   \;\; , \;\;  \langle \phi'\rangle = 1.1 \; {\rm TeV} \; \Rightarrow  \; M_\varphi = 1.54 \; {\rm TeV} ,  \;  m_\psi = 1.1   \; {\rm TeV}   ~~,
 \nonumber \\ [2mm]
&&   M_\Theta = 1.6   \; {\rm TeV}    \;\; , \;\;    \kappa =  1  ~~.
\label{eq:bench2}
\eear
In the case of benchmark point 1, $ \Gamma (A \to A'gg) \approx 40$ eV is more than two orders of magnitude smaller than $\Gamma (A \to \gamma\gamma) \approx 14 $ keV.
The branching fractions of $A$ into $\gamma\gamma$, $Z\gamma$ and  $ZZ$ are thus 
59.1\%, 35.4\% and  5.3\%, respectively.  
For benchmark point 2, $ \Gamma (A \to A'gg) \approx 0.3$ keV and  $\Gamma (A \to \gamma\gamma) \approx 43 $ keV, so 
the $A \to \gamma\gamma$, $Z\gamma$ and  $ZZ$ branching fractions are only slightly smaller than for benchmark point 1.

\subsection{LHC signal rate in the 2-step model}
\label{subsec:lhc2step}

For a $\Theta^a$ mass $M_\Theta >  M_\varphi/2$, the coupling (\ref{eq:kappa}) induces an
interaction of $\varphi$ with gluons approximately given by the dimension-5 operator 
\be
\frac{\alpha_s \kappa\, \langle \phi' \rangle \,  {\cal C}_\Theta {\cal C}_{\rm loop} }{16\sqrt{2} \pi M_\Theta^2  } 
 \; \varphi  \, G^{\mu\nu \, a} \,  G_{\mu\nu}^{\, a}  ~~,
\label{eq:dim5}
\ee
where ${\cal C}_\Theta$ is a coefficient that includes the deviations from the small $M_\varphi^2/(2M_\Theta)^2$ limit 
(the full expression without taking the large $M_\Theta$ limit can be extracted from \cite{Dobrescu:2011aa}): 
\be
{\cal C}_\Theta = 1+ \frac{2 M_\varphi^2}{15 M_\Theta^2} + \frac{3 M_\varphi^4}{140 M_\Theta^4} + O \left( M_\varphi^6/(2M_\Theta)^6 \right) ~~.
\ee
The coefficient ${\cal C}_{\rm loop}$  includes higher-order loop corrections:
\be
{\cal C}_{\rm loop} \simeq 1+ \frac{33\alpha_s}{4\pi} + \frac{5 \lambda_\Theta}{16 \pi^2}   ~~.
\ee
The first term here arises from integrating $\Theta^a$ out at one loop, while the next two terms arise at two loops and have been computed in \cite{Boughezal:2010ry}.
The term proportional to $\lambda_\Theta$ involves one insertion of the quartic $\Theta$ coupling, see Eq.~(\ref{Vtheta}).
We assumed $\mu_\Theta \ll M_\Theta$, so that the 2-loop contributions with trilinear $\Theta$ couplings are negligible.

The dimension-5 operator (\ref{eq:dim5}), which is responsible for $\varphi$ production, also leads to the decays of $\varphi$ into gluons and, at higher-orders in the QCD coupling, into quark pairs.
These decays have a width given at the next-to-leading order by
\bear
&& \hspace*{-2cm} \Gamma (\varphi  \to \! g g, 3g, g q\bar q  ) \simeq \alpha_s^2(\mu_{\rm decay}) \,  \frac{\kappa^2  \langle \phi' \rangle^2  M_\varphi^3 }{256 \pi^3\, M_\Theta^4}  
\,  \, {\cal C}_\Theta^2 \,  {\cal C}_{\rm loop}^2 
 \nonumber \\ [3mm]
 && \hspace*{2.4cm}  \times \left[ 1+ \frac{\alpha_s}{\pi} \left( \frac{73}{4} - \frac{7 N_f}{6} - \frac{33 - 2N_f}{3} \ln \frac{M_\varphi}{\mu_{\rm decay}} \right) \! \right]  ~~.
\label{eq:gg}
\eear
The next-to-leading order corrections \cite{Djouadi:2005gi}  shown here depend on the number $N_f$ of quark flavors lighter than $M_\varphi/2$. 
As the $g t\bar t $ final state is phase-space suppressed, the effective value of $N_f$ is between 5 and 6.
These corrections also depend on a renormalization scale, which is taken to be 
 $\mu_{\rm decay} = M_\varphi$ in order to minimize higher-order corrections to the decay width \cite{Steinhauser:1998cm}. 
The QCD coupling constant decreases from  $\alpha_s (M_\varphi) \approx 0.092$ at the scale  $M_\varphi = 780$ GeV to 
$\alpha_s (M_\varphi) \approx 0.085$ at $M_\varphi = 1.54$ TeV \cite{Agashe:2014kda}.

Besides the unavoidable decay into jets, there are a few other more model-dependent decay modes.
Notably, the $\phi$ and $\phi'$ fields can couple to the SM Higgs doublet, $H$, via $|\phi|^2 H^\dagger H$, $|\phi'|^2H^\dagger H$,  and $\phi \phi' H^\dagger H$ terms.
The latter, in particular, must have a suppressed coefficient (below $\sim 3 \times 10^{-2}$) to avoid a large mixing of $\varphi$ with the SM Higgs boson $h^0$. Otherwise, the dominant 
decay modes of $\varphi$ would be into $WW, ZZ$ and $t\bar{t}$, and the branching fraction $B(\varphi \to A A')$ would be too small to yield an observable diphoton signal. We ignore the $\varphi - h^0$ mixing in what follows.

In addition, the small but nonzero $\langle \phi \rangle$ discussed in \Sref{subsec:scalarints} leads to 
the $\varphi \to A'A'$ decay.  The width for this decay is not phase-space suppressed, but it is proportional to 
$(\langle \phi \rangle/\langle \phi' \rangle)^2 < 10^{-2}$, and it can be neglected compared to 
the $\varphi \to A'A$ width.
Another subdominant decay of $\varphi$ is into $\gamma\gamma$, due to a $\psi$ loop; its branching fraction, of order 0.1\%, is too small to be phenomenologically relevant. 

Comparing the main $\varphi$ decay widths given in Eqs.~(\ref{eq:phiAA}) and (\ref{eq:gg}), we find that 
the branching fraction for $\varphi \to A A'$ is sizable for a large range of parameters. 
For example, the benchmark point 1 [see Eq.~(\ref{eq:bench})] implies $B(\varphi \to A A') \approx 91\%$ and a total  width for 
$\varphi$ given by  $\Gamma_\varphi  \approx 6 \times 10^{-6} M_\varphi$, while benchmark point 2 [see Eq.~(\ref{eq:bench2})]  gives $B(\varphi \to A A') \approx 70\%$ and 
$\Gamma_\varphi  \approx 4 \times 10^{-6} M_\varphi$.

The dimension-5 operator (\ref{eq:dim5}) also leads to $s$-channel production of $\varphi$ at the LHC. 
In that case, the renormalization scale that minimizes the higher-order corrections is approximately $M_\varphi/2$.
Including the interaction (\ref{eq:dim5}) in FeynRules \cite{Alloul:2013bka} with a coefficient that depends on the QCD coupling constant  at the $M_\varphi/2$ scale, $\alpha_s(M_\varphi/2)$,
we have generated the model files for 
MadGraph \cite{Alwall:2014hca} and obtained the leading-order cross section for inclusive $\varphi$ production at the 13 TeV LHC, $\sigma_{\rm LO} (p p \to \varphi X)$.

Let us focus first on benchmark point 1, for which we calculate
\be
\sigma_{\rm LO} (p p \to \varphi X) \approx 6.8 \; {\rm fb} \, \,  
  \left(\frac{\langle \phi'\rangle}{680 \; {\rm GeV} } \right)^{\! 2}   \kappa^2
\ee
for $M_\varphi = 780$ GeV [corresponding to $\alpha_s(390 \; {\rm GeV}) \approx 0.097$], $M_\Theta = 800$ GeV and, less importantly, $\lambda_\Theta =1$. Recall that Eq.~(\ref{eq:lambda}) implies 
$\langle \phi'\rangle = 680$ GeV for $\lambda = 0.1$. 

Higher-order corrections to $\varphi$ production are large; we break them down as follows: \\ [1.1mm]
1) The next-to-leading order QCD corrections, using the dimension-5 interaction (\ref{eq:dim5}), which include 1-loop corrections as well as a real emission from the initial state partons.
We have computed those using the MCFM code \cite{Campbell:2015qma}, and obtained a multiplicative $K$ factor given by $K_{\rm NLO} = 1.88$ for $M_\varphi = 780$ GeV. \\  [1.5mm]
2) The NNLO and N$^3$LO QCD corrections, again in the large $M_\Theta$ limit.
These have been computed in the case of Higgs production \cite{Anastasiou:2016cez}, and amount to an additional 30\% increase.
In the case of our $\varphi$ production, 
we expect that these corrections are smaller by a factor of roughly $\alpha_s(M_\varphi/2)/ \alpha_s(M_h/2) \approx 0.8$ for $M_\varphi = 780$ GeV, so that the total  multiplicative factor becomes
$K_{\rm N^3LO} \sim 1.24 K_{\rm NLO} \approx 2.3$. This estimate is consistent with the recent
result of Ref.~\cite{Anastasiou:2016hlm} ($K_{\rm N^3LO} \approx 2.3$ for a scalar of mass at 750 GeV). \\  [1.5mm]
3) Finite $M_\Theta$ effects on the QCD corrections. We will neglect these here.  \\ [1.5mm]
Although the higher-order corrections to $\varphi$ production are smaller than the ones to Higgs production in the SM, they are essential for computing the correct signal rate.

\begin{figure}[t] 
   \centering
   \hspace*{-1mm} \includegraphics[width=0.47\textwidth, angle=0]{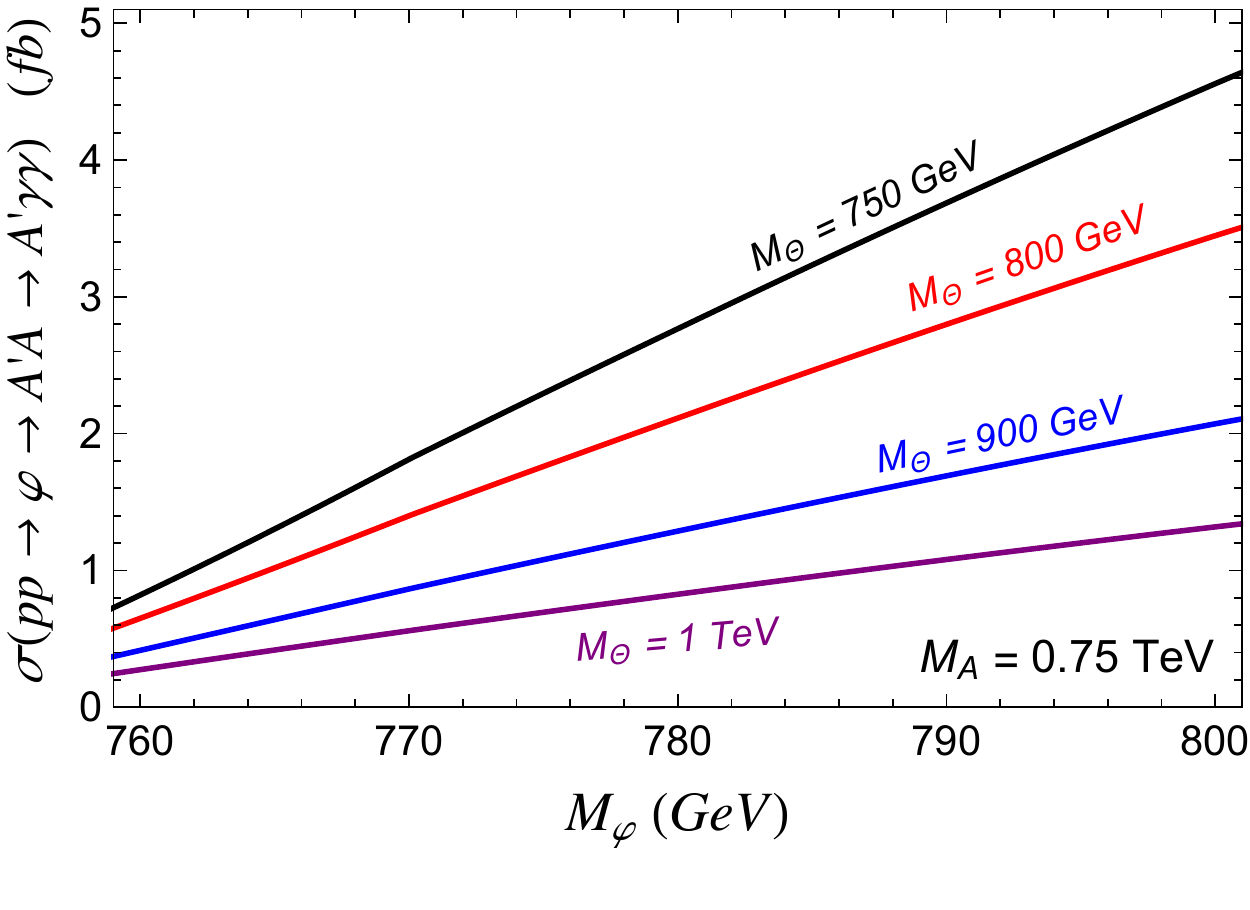} \  \includegraphics[width=0.5\textwidth, angle=0]{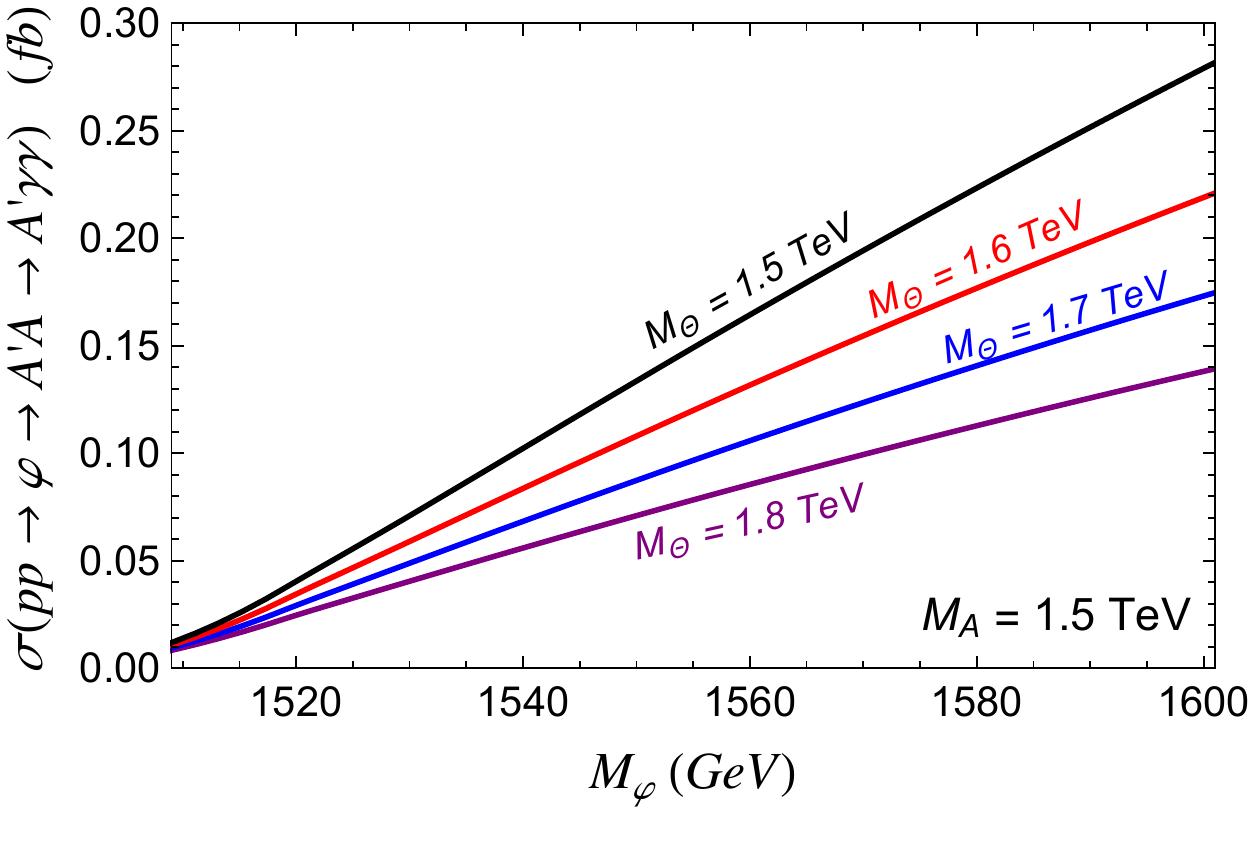}
   \caption{Diphoton signal cross section at the 13 TeV LHC in the 2-step model, as a function of the $\varphi$ mass, for four different masses of the color-octet scalar $\Theta$ responsible for gluon fusion.
 Left  panel corresponds to $M_A = 750$ GeV and  $\kappa = 0.5$, and right panel corresponds to $M_A = 1.5$ TeV and  $\kappa = 1$ (the cross section scales as $\kappa^2$).
  The other parameters used here are $\lambda = 0.1$, $\lambda_\Theta = y_\psi' = 1$ and $M_{A'} = 0$.}
   \label{fig:xsecMphi}
\end{figure}

The $s$-channel production of $\varphi$ is followed by the $\varphi \to A A'$ and $A\to  \gamma\gamma$ decays (see \Fref{fig:phiAdiagram}).
The nonresonant contribution to $gg \to A A'$  due to the (\ref{eq:AAgg}) interaction is two orders of magnitude smaller, and it can be safely ignored.
The total cross section times branching fractions of a diphoton signal  at the 13 TeV LHC is thus
\bear
&&  \sigma_{\gamma\gamma}  \equiv \sigma (p p \to \varphi \to A' A \to A' \gamma\gamma)
\nonumber \\ [3mm]
&& \hspace*{0.7cm}
 = K_{\rm N^3LO} \; \sigma_{\rm LO} (p p \to \varphi X) \, B(\varphi \to A A')  \,  B(A \to \gamma\gamma)  ~~.
\eear
For $M_\varphi = 780$ GeV, $M_\Theta = 800$ GeV, $M_A = 750 $ GeV, $M_{A'} \ll M_\varphi - M_A$,  $\kappa = 0.5$
and using the values for dimensionless couplings given in Eq.~(\ref{eq:coupling-values}), we find $ \sigma_{\gamma\gamma} \approx 2.1$ fb.
For other values of $M_\varphi$ and $M_\Theta$, the signal  cross section $\sigma_{\gamma\gamma}$ is shown in the left panel of \Fref{fig:xsecMphi}.

In the case of benchmark point 2 [see Eq.~(\ref{eq:bench2})], where $M_\varphi = 1540$ GeV, $M_\Theta = 1.6$ TeV, $M_A = 1.5 $ TeV, $M_{A'} \ll M_\varphi - M_A$ and  $\kappa = 1$,
we find $K_{\rm N^3LO} \approx 2.2$ and  $ \sigma_{\gamma\gamma} \approx 0.08$ fb.
Varying  $M_\varphi$ and $M_\Theta$ while keeping the other parameters fixed gives the values of $\sigma_{\gamma\gamma}$ shown in the right panel of \Fref{fig:xsecMphi}.
Thus, it is clear that the 2-step model may lead to signal rates that are large enough to be observed at the 13 TeV LHC even for a diphoton resonance as heavy as 1.5 TeV.

Let us comment on limits relevant for the other particles present in this model.
The current lower 
limit on $M_\Theta$ can be derived from the CMS search in the final state with a pair of dijet resonances of equal mass \cite{Khachatryan:2014lpa}.
The cross section for the process $pp \to \Theta \Theta$ depends on a single parameter, $M_\Theta$, as the $\Theta$ interaction with gluons is fixed by QCD,
and there is no $SU(2)_W$ invariant coupling of  $ \Theta$ to the SM quarks. 
The 1-loop process $\Theta \to gg$, which proceeds through the trilinear coupling $\lambda_\Theta$, has a branching fraction near 100\%.
The CMS limit of about 1 pb on the cross section for pair production of dijets corresponds to $M_\Theta \gsim 400$ GeV 
(note that the theoretical prediction shown in Figure 7 of  \cite{Khachatryan:2014lpa} refers to a spin-1 particle, whose production rate is almost 40 times larger than 
for the spin-0 color octet discussed here \cite{Dobrescu:2007yp}).

In order to avoid observable $A'$ decays into photons, we 
take the $A'$ mass $M_{A'}$ sufficiently small so that the decay is outside the electromagnetic calorimeter. 
The decay length of a light $A'$ in the rest frame is 
\be
L_{A'} \approx  \frac{y^2_\psi M_A^3}{y^{\prime 2}_\psi M_{A'}^3 \,  \Gamma (A \to \gamma\gamma)}  ~~,
\ee
where $y_\psi'$ is related to $m_\psi$ by Eq.~(\ref{eq:mpsi}).
In the lab frame, the decay length is increased by $E_{A'}/M_{A'}$, where the $A'$ energy $E_{A'}$ is of the order of $M_\varphi - M_A$.
For example, if $M_{A'} < 0.4$ GeV, $y_\psi'=y_\psi$, and $M_\varphi = 780$ GeV, then the $A'$ decay length is longer than 7 m.

A lower limit on $M_{A'}$ is set by star cooling constraints. Values of
$M_{A'}$ below about 6 MeV make the decay length comparable to the size of a supernova core, so $A'$ emission may modify 
the supernova temperature \cite{Raffelt:1996wa}.  
Even for $M_{A'} < 6$ MeV though, the $A'$ mean-free path may be smaller than the supernova core because an 
$A'A' gg$ interaction is induced at one loop by the potential term proportional to $\kappa_2$ in Eq.~(\ref{eq:Vtheta}). In any case,
the range of $M_{A'}$ consistent with all the constraints spans at least two orders of magnitude.
Thus, we can assume that  $M_{A'} \ll M_\varphi - M_A$, and that the  $A'$ gives rise to missing transverse energy at the LHC.

The vectorlike lepton $\psi$ is a weak singlet and has hypercharge +1 ({\it i.e.}, electric charge +1). Therefore,
an $H \bar L_L \psi_R$ Yukawa coupling to the SM lepton doublets $L$ is gauge invariant.  This coupling leads to mixing 
of $\psi$ with the SM charged leptons. As a result, the main decay modes of the new heavy fermion are $\psi \to W\nu, Z\tau, h^0 \tau$.
The lower limits on $m_\psi$ set at colliders are loose, of order 100 GeV \cite{Falkowski:2013jya}. 

Searches for vectorlike leptons produced in pairs at the LHC will provide a test for this model. One should recognize though that there is 
some flexibility in choosing the particles running in the loops that lead to $A \to \gamma\gamma$. For example, if instead of a vectorlike lepton
there is a charged scalar that couples to $\phi$, then the diphoton signal would not changed. By contrast, the presence of the color-octet 
scalar $\Theta$ in the production loop is a more robust feature. Note that if instead of $\Theta$ there were a vectorlike quark
responsible for $\varphi$ production through gluon fusion, then it would be difficult to avoid the coupling of $A$ to the vectorlike quark so that the main
decay of  $A$  would be into gluons, rendering a too small rate for $A \to \gamma\gamma$. 
Thus, searches for pair production of $\Theta$ are a more generic test  of this 2-step model.

\bigskip

\section{A strange-production model} \setcounter{equation}{0}
\label{sec:3step}

In this section, we construct a model in which a diphoton resonance arises as part of a cascade decay of a new gauge boson $Z'$ 
produced predominantly via strange-quark fusion
\be
p p \rightarrow Z' \rightarrow \bar \nu_s  N  \; , \; \; N \rightarrow \nu_s \Phi \; , \; \; \Phi \rightarrow \gamma \gamma
\ee
where $\Phi$ is a (pseudo)scalar with a mass $M_\Phi$ near the TeV scale, $N$ is a SM-singlet heavy Dirac fermion, and $\nu_s$ is a sterile neutrino. 
Unlike the gluon-initiated model of \Sref{sec:2step}, this model relies on sea-quark production. The production process could potentially be distinguished using differential distributions. For instance, variation in QCD radiation between quarks and gluons could lead to discernible differences in $N_{\rm jet}$ distributions, as we will show in \Sref{sec:distributions}.

\subsection{New fields and symmetries}

We introduce a new gauge symmetry $U(1)_{sb}$ under which the only SM states that are charged are the oppositely-charged right-handed $s$ and $b$ quarks.  The  $Z'$ couplings are flavor diagonal as long as the right-handed quark mass and gauge eigenstates coincide, and so are not subject to constraints from flavor processes (\eg, $B_s$ mixing).
This charge assignment leads to a $U(1)_Y \left[U(1)_{sb} \right]^2$ anomaly, requiring the introduction of new fermions $f, f'$ (``anomalons'') that are vectorlike with respect to the SM gauge groups (and $SU(3)_c \times SU(2)_W$ singlets) but chiral under $U(1)_{sb}$.
An implication of the anomalons charged under both $U(1)_Y$ and $U(1)_{sb}$ is the loop-induced decays of the new (pseudo)scalars to two photons.\footnote{A model with similar charge assignment was proposed in a completely different context in \cite{Dobrescu:2014fca}.  As an alternative model, the coupling of the $Z'$ to the SM could be through a higher-dimension operator, generated by integrating out fermions transforming under the SM gauge group as the right-handed strange quark \cite{Fox:2011qd}. } 
We include three Weyl fermions $N_{\pm R}, N_L$, which are
SM singlets and permit the cascade decay of the $Z'$.
The relevant matter content of the theory is given in Table~\ref{table:U1}.

\begin{table}
\renewcommand{\arraystretch}{1.5}
\centering\begin{tabular}{M | M M M | M}\hline 
\text{field}  & SU(3)_c  & SU(2)_W   &  U(1)_Y  & \  U(1)_{sb}  \\  [-0.03em] \hline\hline
s_R & 3 & \  1 \ & \  \!\!-1/3  &   +1  \\ [0.1em]  
b_R  & 3 & \  1 \ & \  \!\!-1/3  &   -1  \\ [0.1em]  
\hline
f_R   & 1 &      1      &   \!\!+1    &    +1  \\ [0.1em]      
f_R'  & 1 &   1      &   \!\!+1       &    -1  \\ [0.1em]      
f_L, f_L'   &  1 &      1     &   \!\!+1     &   0  \\ [0.1em]      
\hline
{N_+}_R  &  1 &      1     &   0     &   +1  \\ [0.1em]      
{N_-}_R   &  1 &      1     &   0     &   -1  \\ [0.1em]      
N_L   & 1 &      1      &   0    &    0  \\ [0.1em]      
\hline
\phi' , \phi & 1 & \  1 \ & \  0 \  &   +1    \\ [0.3em]   
 \hline
\end{tabular}
\medskip \\
\caption{\small Fields charged under the $U(1)_{sb}$ gauge symmetry. The spin is 0 for $\phi'$ and $\phi$, and 1/2 for 
 the anomalons ($f$, $f'$), the $N$  fields, and the $s_R$ and $b_R$ quarks. }
\label{table:U1}
\end{table}

Along with the new fermions, the theory contains two new scalars charged under $U(1)_{sb}$, $\phi'$ and $\phi$. We assume to a first approximation (discussed later) that the scalar potential respects a ${\mathbb Z}_2$ symmetry under which $\phi \rightarrow - \phi$ and $\phi' \rightarrow \phi'$.  Constraints on deviations from SM Higgs branching fractions, as well as the dihiggs production rate and $t\bar{t}$ resonance searches limit the size of couplings between the SM Higgs and $\phi$ and $\phi'$ so we also assume that the couplings to the SM Higgs field are small (we will discuss this more in \Sref{sec:productionanddecay}).  

Under these assumptions, the most general renormalizable potential for the new spin-0 fields is
\be
V \supset - M_{\phi'}^2 \abs{\phi'}^2 + M_{\phi}^2 \abs{\phi}^2 + \frac{\lambda_{\phi'}}{4} \abs{\phi'}^4  +  \frac{\lambda_{\phi}}{4} \abs{\phi}^4 + \lambda_{\phi' \phi} \abs{\phi'}^2 \abs{\phi}^2 + \left[ \lambda'_{\phi' \phi} (\phi^\dagger\phi')^2 +\mathrm{H.c.} \right]~~.
\label{eq:threestepscalarpot}\ee
We take the quartic couplings and the mass-squared parameters to be positive. In this limit, a non-zero VEV develops for $\phi'$ while $\vev{\phi} = 0$. As such, $\phi'$ is responsible for spontaneously breaking $U(1)_{sb}$, giving mass to the $Z'$ as well as to the anomalons.
The additional scalar $\phi$ will be the diphoton resonance, as we discuss below.  After $U(1)_{sb}$ breaking the real spin-0 fields are
\be
\phi' = \vev{\phi'} + \frac{\varphi'}{\sqrt{2}} \quad \quad, \quad \quad \quad \phi = \frac{1}{\sqrt{2}} \left(\varphi + i A\right).
\ee
The mass of the new gauge boson $M_{Z'} = \sqrt{2} g_z \vev{\phi'}$, where $g_z$ is the $U(1)_{sb}$ gauge coupling.  We assume that $M_{\varphi'}$ is sufficiently large that this state does not play a role in the phenomenology of interest.

The scalars couple to anomalons via
\be
{\cal L} \supset - y'_f \phi'^\dagger \bar{f}_L f_R - y'_{f'} \phi' \bar{f}'_L f'_R - y_f \phi^\dagger \bar{f}_L f_R - y_{f'} \phi \bar{f}'_L f'_R + \mathrm{H.c.}
\ee
The anomalons acquire a mass from their coupling to $\phi'$, $m_{f^{(\prime)}} = y'_{f^{(\prime)}} \vev{\phi'}$, while the Yukawa couplings of $\phi$
permit the physical states $\varphi$ and $A$ to decay to photons at one loop.
We impose $m_f, m_{f'} > M_{Z'}/2 $ to forbid the decay of $Z'$ into anomalons.
Additional Yukawa terms of the type $\phi'^\dagger \bar f'_L f_R$ or $\phi' \bar f_L f_R^\prime$, or similarly with 
$\phi'$ replaced by $\phi$, may be present, 
leading to $f-f'$ mixing; however, such mixing is not consequential in what follows. 

The SM Yukawas for $b$ and $s$ quarks are forbidden by $U(1)_{sb}$ but are allowed at dimension 5 and can be generated from a renormalizable theory by integrating out vectorlike fermions that have the same quantum numbers as $b_R$.  For order one couplings these heavy fermions must be lighter than $\langle\phi'\rangle/y_q^{SM}\sim 10-100\ \mathrm{TeV}$.  There are additional renormalizable terms $\phi' \bar{f}^c_R \ell_R$ which mix the anomalons with the right-handed leptons and allow the anomalons to decay to $W\nu,\,Z\ell$.

The scalars also couple to the $N$ fermions:
\be
{\cal L} \supset - y'_+ \phi'^\dagger \bar{N}_L {N_+}_R - y'_- \phi' \bar{N}_L {N_-}_R - y_+ \phi^\dagger \bar{N}_L {N_+}_R - y_- \phi \bar{N}_L {N_-}_R +  \mathrm{H.c.}
\label{eq:ypmyukawas}
\ee
We assume that the $N$-number violating mass terms of the form $\bar{N}_L N_L^c$ and $\bar{N}^c_{+\, R} {N_-}_R$ are suppressed (or even completely forbidden) by a global $U(1)_N$ symmetry.
As such, for $\vev{\phi'} \neq 0$, the couplings $y'_\pm$ give mass to a Dirac fermion $N$, whose left-handed component is 
$N_L$ (see Table 1), and whose right-handed component is a linear combination of ${N_\pm}_R$:
\be
N_R = c_N {N_+}_R + s_N {N_-}_R  ~~.
\ee
The orthogonal linear combination of ${N_\pm}_R$ forms a massless 2-component fermion $\nu_s$, 
\be
 {\nu_s}_R  = - s_N {N_+}_R + c_N {N_-}_R  ~~,
\ee
with the mixing given by
\be
s_N , c_N = \frac{y'_\mp}{\sqrt{{y'_+}^2 + {y'_-}^2}}
~~.
\ee
As $\nu_s$ is a massless (or nearly massless if a small Majorana mass is introduced) fermion that is a SM singlet, it is appropriate to refer to it as 
a sterile neutrino. Higher-dimensional operators may induce some small mixing between $\nu_s$ and the SM neutrinos; we will ignore here these effects.
The mass of $N$ is related to the $\phi'$ VEV by
\be
m_N = \frac{y'_-}{s_N} \vev{\phi'} ~~.
\ee

\begin{figure}[b] 
\vspace*{-2.cm}
   \centering
    \includegraphics[width=1\textwidth, angle=0]{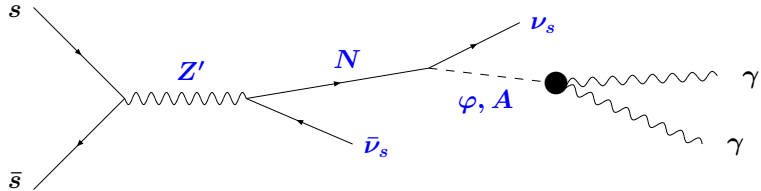}
\vspace*{-7.5cm}
   \caption{Diagram for $\varphi/A$ production at the LHC in the 3-step $Z'_{sb}$ model. The sterile neutrinos $\nu_s$ escape the detector. }
   \label{fig:phidiagrams}
\end{figure}

 The Yukawa couplings of the components of $\phi$ to the physical fermions $N$ and $\nu_s$ are 
\be
{\cal L} \supset -  \frac{1}{\sqrt{2}} \Big[ (- y_+ s_N + y_- c_N) \varphi + (y_+ s_N + y_- c_N) i A \Big] \bar{N}_L {\nu_s}_R + 
\mathrm{H.c.} 
\ee
Note that any couplings of $\varphi$ or $A$ that can lead to their decays to $\nu_s\nu_s$ vanish in the limit of massless $\nu_s$.  
The couplings of the $Z'$ boson to the electrically-neutral fermions are given by
\be
{\cal L}_{\rm kin} \supset g_z (c_N^2 - s_N^2) \left(\bar{N}_R \slashed{Z}' N_R - {\overline\nu_s}_R \slashed{Z}'  {\nu_s}_R \right) 
- 2 g_z c_N s_N \left(\bar{N}_R \slashed{Z}' {\nu_s}_R  + \mathrm{H.c.}  \right)~~.
\ee
We now have the interactions necessary to describe the production and decay of the particles that lead to a diphoton signal, as shown in \Fref{fig:phidiagrams}.  Note that, unlike the previous model, there is no symmetry reason why the masses of $Z'$, $N$, and the scalar leading to the diphoton resonance should all be similar; although their masses are proportional to the same VEV, they involve unrelated couplings.  However, it is possible that these couplings are related through renormalization group evolution \cite{Kearney:2013xwa}.

\subsection{Production and decay}\label{sec:productionanddecay}

The dominant partial width of the $Z'$ is to pairs of right-handed $s$ and $b$ quarks,
\be
\Gamma(Z'\rightarrow s\bar{s}) \simeq \Gamma(Z'\rightarrow b\bar{b}) \simeq \frac{g_z^2 }{8\pi}  M_{Z'} ~~,
\ee
while the widths to the neutral fermions are
\bea
\Gamma(Z'\rightarrow \nu_s \bar \nu_s) &=& \frac{g_z^2 (c_{N}^2 - s_N^2)^2}{24\pi}  M_{Z'} ~~,  \nonumber \\ [2mm]
\Gamma(Z'\rightarrow N \, \nu_s) &=& \frac{g_z^2 s_{N}^2 c_N^2 }{3\pi} M_{Z'} \left(1-\frac{3m_{N}^2}{2M_{Z'}^2}+\frac{m_{N}^6}{2M_{Z'}^6}\right)~~.
\eea
In a compressed spectrum the second decay is phase-space suppressed, $\Gamma(Z'\rightarrow N  \, \nu_s) \approx g_z^2 s_{N}^2 c_N^2 (M_{Z'}-m_{N})^2/ (\pi M_{Z'})$.  

In turn, the 2-body decays of $N$ are
\bea
\Gamma(N\rightarrow \varphi \, \nu_s) &=& \frac{(- y_+ s_N + y_- c_N)^2 }{64\pi} \,  m_{N} \left(1-\frac{M_\varphi^2}{m_{N}^2}\right)^{\! 2} ~~, \nonumber \\ [2mm]
\Gamma(N\rightarrow A \, \nu_s) &=& \frac{(y_+ s_N + y_- c_N)^2}{64\pi} \, m_{N} \left(1-\frac{M_{A}^2}{m_{N}^2}\right)^{\! 2}~~.
\eea
These also become small as the splitting between $N$ and the scalar in the final state becomes small.  Thus, the 3-body decay to a pair of $s$ or $b$ quarks, through an off-shell $Z'$, may compete.  This width is
\be
\Gamma (N\rightarrow \nu_s q\bar{q})  = \frac{3g_z^4 \, s_{N}^2 c_N^2 \, M_{Z'}^2}{16\pi^3 m_{N}}\left(1-\frac{m_{N}^2}{2M_{Z'}^2}-\frac{m_{N}^4}{6M_{Z'}^4}-\left(1-\frac{M_{Z'}^2}{m_{N}^2}\right)\ln \left(1-\frac{m_{N}^2}{M_{Z'}^2}\right)\right)~~,
\ee 
where we have ignored the quark masses. 
The missing energy is difficult to observe at the LHC if the mass splittings $M_{Z'} \gsim m_{N} \gsim M_A$ are small. As a result of these small splittings, $N$ decays to the diphoton resonance are phase-space suppressed, while the 3-body decays to $\nu_s q\bar{q}$ are not.
To compensate for this and achieve a significant diphoton rate the couplings $y_\pm$ cannot be too small. 
Furthermore, the phase-space suppression for $Z'\rightarrow N  \nu_s$ means that there can be substantial contribution to $N \nu_s$ production from an off-shell $Z'$. This non-resonant production of the diphoton state can alter the kinematic distributions considerably, injecting additional missing transverse energy into the event relative to the case of on-shell production.   

Note that this feature does not occur in the 2-step model, but rather represents a particular challenge for a vector cascade model due to the phase-space suppressed production of fermions via a vector boson. In the center-of-mass frame, the 3-step process is proportional to the final state velocity $\beta_f^3$, whereas near-threshold scalar production from a scalar resonance (as in the 2-step model) is proportional to $\beta_f$. As a result, gains from going off-shell are more substantial in the 3-step model, leading to a significant contribution from non-resonant $Z'$ production.

So far, we have remained agnostic as to whether $\varphi$ or $A$ is the observed resonance, since both can decay to diphotons (as well as $ZZ$ and $\gamma Z$) through a loop of anomalons.  Integrating out the anomalons leads to effective dimension-5 operators coupling the scalars to the hypercharge field strength,
\be
\frac{\alpha}{6\sqrt{2}\pi\cos^2\theta_w}\left(\frac{y_f}{m_f} + \frac{y_{f'}}{m_{f'}}\right) \varphi B^{\mu\nu}B_{\mu\nu} 
+ \frac{\alpha}{8\sqrt{2}\pi\cos^2\theta_w} \left(-\frac{y_f}{m_f} + \frac{y_{f'}}{m_{f'}}\right) A \, B^{\mu\nu}\tilde{B}_{\mu\nu} ~~.
\ee
The resulting widths are
\bea
\Gamma(\varphi \rightarrow \gamma\gamma) &=& \frac{\alpha^2}{288\pi^3}\left(\frac{y_f}{m_f} + \frac{y_{f'}}{m_{f'}}\right)^2 M_\varphi^3  ~~, \\
\Gamma(A \rightarrow \gamma\gamma) &=& \frac{\alpha^2}{128\pi^3}\left(-\frac{y_f}{m_f} + \frac{y_{f'}}{m_{f'}}\right)^2 M_{A}^3~~.
\eea
As the anomalons are $SU(2)_W$ singlets, the $A$ pseudoscalar exhibits characteristic branching fractions to diphotons
$B(A \rightarrow \gamma \gamma) = 59.5\%$,
and so can serve as the resonance.  Whether or not $\varphi$ has the same branching fraction and also contributes to the diphoton rate depends on details of the model we have not yet discussed.  Specifically, additional terms in the scalar potential can give rise to terms that mix $\varphi$ and the SM Higgs boson.\footnote{Terms mixing $\varphi'$ and $h^0$ may also be present, but are irrelevant for the diphoton signal. Such terms must simply be small enough to be consistent with measurements of Higgs couplings.}
For instance, a term
\be
\Delta V = \lambda_{\phi \phi' H} \left(\phi^\dagger \phi' + \phi \phi'^\dagger\right) \abs{H}^2
\ee
will lead to $\varphi-h^0$ mixing, as well as inducing a VEV for $\phi$. Note that this term violates the ${\mathbb Z}_2$ symmetry $\phi \rightarrow -\phi$, but this parity is also collectively broken by the Yukawas, so it cannot be used to set the above term to zero.
The explicit breaking by the Yukawa couplings to the anomalons leads to this term being generated at two loops, so it is consistent to treat $\Delta V$ as a small perturbation on the potential of (\ref{eq:threestepscalarpot}).\footnote{Similar reasoning motivates neglecting the mass-mixing term $b^2 \phi^\dagger \phi'$, which only arises at one loop.} 

Thus, one can imagine two scenarios.  In the first case, the $\varphi-h^0$ mixing is very suppressed, $\lambda_{\phi \phi' H} \lsim 10^{-4}$, such that $\varphi$ decays predominantly via anomalon loops. Then, both $A$ and $\varphi$ can contribute to the diphoton rate.  
A mass splitting between $\varphi$ and $A$ could broaden the diphoton signal, potentially leading to the intepretation that the signal arises from a single, wide resonance. This splitting can be generated from $\Delta V$,  $\lambda'_{\phi' \phi}$ and the mass-mixing term.

Alternatively, $\varphi-h^0$ mixing may be non-negligible, leading to additional decays of $\varphi$ to SM states, notably $\varphi \rightarrow WW, ZZ, t \bar{t}$. As such, the diphoton signal comes entirely from $A$ and the model predicts a second resonance of mass close to $M_A$ with branching fractions characteristic of a singlet mixing with the Higgs.  Since $\varphi$ no longer contributes to the diphoton rate, the decay $N \rightarrow A \, \nu_s$ must dominate over $N\rightarrow \varphi \, \nu_s$, for instance due to $M_\varphi > m_{N}$ or coincidental cancellations as for $y_+ \simeq y_-$ (supposing $c_N \simeq s_N$).

We now present a example parameter point with a diphoton resonance at 750 GeV:
\bea
&& M_{Z'} = 790\, \GeV \;\;  ,  \;\; \, g_z=0.3  \, \Rightarrow \, \langle \phi'\rangle \approx 1.9 \TeV~~, \nonumber\\
&& m_{N} = 760\, \GeV   \;\;  ,  \;\; \, y'_+ =0.3 \, \Rightarrow \, y'_- \approx 0.28~~,\nonumber\\
&& m_f=m_{f'}=500\, \GeV \, \Rightarrow \, y'_f=y'_{f'} \approx 0.27~~.
\label{eq:threestepexampleppoint}
\eea 
At this point the $Z'$ width and relevant branching fraction are $\Gamma_{Z'}\approx 5.7\,\GeV,\, B(Z'\rightarrow N \, \nu_s)\approx 2.7 \times 10^{-3}$.  However, since the phase space for the decay $Z'\rightarrow N \,  \nu_s$ is limited, approximately $20\%$ of $\sigma(pp\rightarrow  N \,  \nu_s)$ comes from off-resonance production.  Including a $K$-factor\footnote{Due to different production channels, the NLO correction for a $Z'$ produced from sea quarks could be up to twice as large as that for a sequential $Z'$, $K_{\rm NLO} \approx 1.2$ \cite{Aad:2014cka}.  
We view our estimate as conservative.} of $K_{\rm  NLO}=1.3$ and using MadGraph \cite{Alwall:2014hca} to calculate the leading order production cross section we find 
\be
\sigma_{\rm 13 TeV}^{\rm NLO}(p p \rightarrow  N \,  \nu_s ) \simeq 5.4 \fb~~.
\ee

\begin{figure}
\centering
\includegraphics[width=0.45\textwidth]{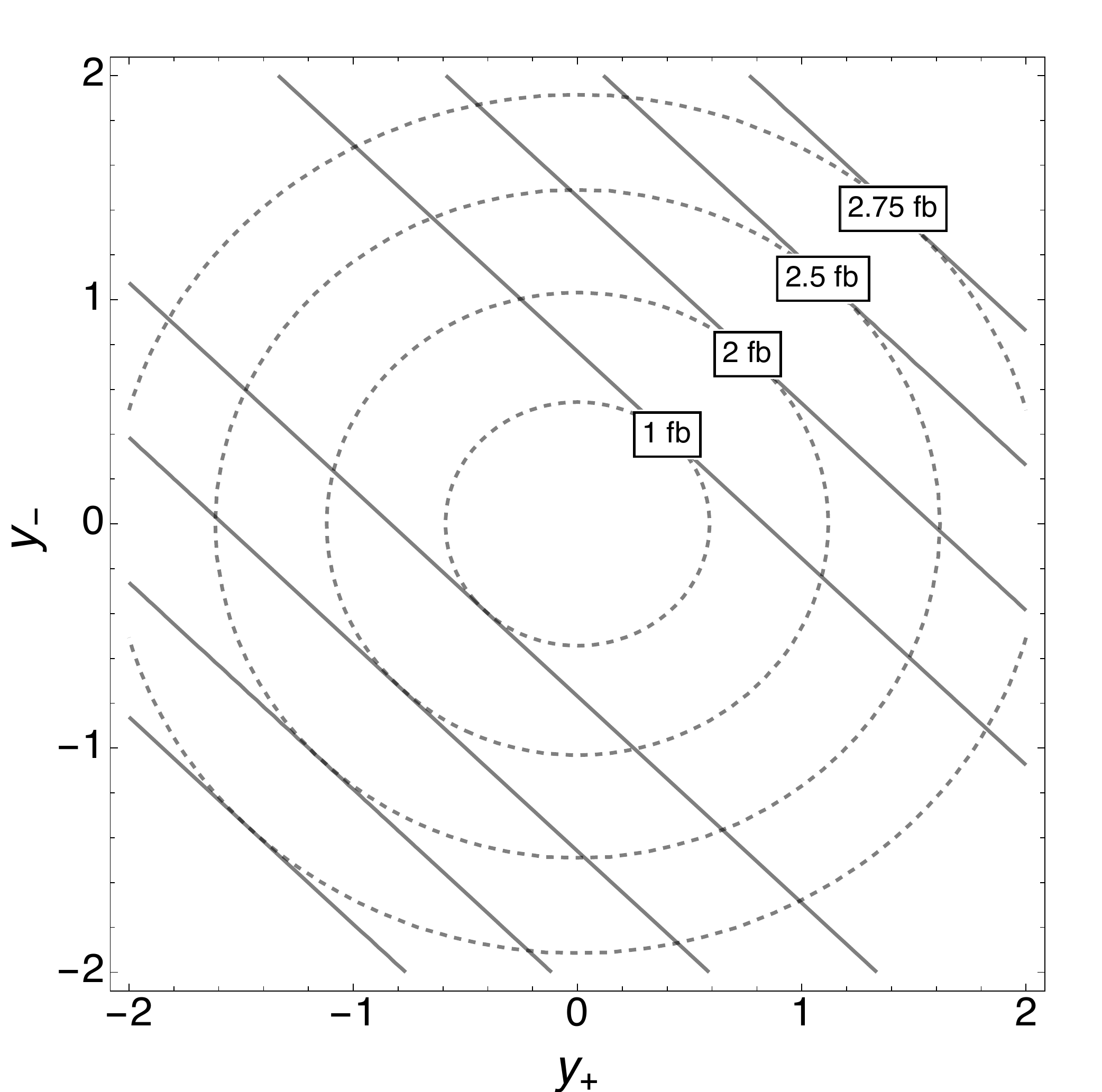}
\caption{Contours of signal cross section $\sigma_{\gamma \gamma}$ as a function of the Yukawa couplings of $\phi$ to $N$, $y_\pm$ (see Eq.~(\ref{eq:ypmyukawas})).  The parameters are as described for the example point (\ref{eq:threestepexampleppoint}).  Solid contours correspond to the case where only the pseudoscalar $A$ contributes to the signal, whereas dashed contours include both decays $N \rightarrow \varphi \, \nu_s , \, A \, \nu_s$.}
\label{fig:ypmp}
\end{figure}

The width for the 3-body decay of $N$ is $\Gamma(N \rightarrow \nu_s j j) \approx 1.9 \MeV$ and the values of $y_\pm$ must be large enough that the 2-body decay $N \rightarrow \nu_s \varphi/A $ dominates this small width.  The required size will depend upon whether both $\varphi$ and $A$ contribute to a $\gamma\gamma$ signal at 750 GeV or just $A$.  In \Fref{fig:ypmp}, we show the signal rate for both possibilities
\be	
\sigma_{\gamma \gamma} = K_{\rm NLO} \, \sigma_{\rm LO}(p p \rightarrow  N \,  \nu_s) \, B(N \rightarrow \Phi \nu_s) \, B(\Phi \rightarrow \gamma \gamma)  ~~,
\ee
where $\Phi$ can represent only $A$ or a combination of $A$ and $\varphi$. We show the corresponding kinematic distributions for this parameter point in \Sref{sec:distributions}.  For this benchmark point the production cross section of $Z'$ at the 8 TeV LHC is, 
\be
\sigma_{\rm 8 TeV}^{\rm NLO}(p p \rightarrow Z' ) \simeq 0.4 \pb~~,
\ee
again including a $K$-factor of $K_{\rm NLO}=1.3$.  This is consistent with the current limits on dijet resonances \cite{Khachatryan:2016ecr}.

We do not consider a second, heavier benchmark for two reasons.  First, the tuning necessary to achieve a compressed spectrum increases with the overall mass scale.  Second, the dijet constraints from the 13 TeV LHC~\cite{CMS:2016wpz} are already significant for $Z'$ mass above $\sim 1$ TeV.  Therefore, it is unlikely that a heavier version of our three-step model would be first observed through final states involving diphotons rather than through $Z'$ decays to dijets.

\bigskip

\section{Kinematic distributions}
 \label{sec:distributions}
 \setcounter{equation}{0}

If a high-mass diphoton resonance will be observed at the LHC, then the 2-step model (presented in \Sref{sec:2step}) and the 3-step model  (presented in \Sref{sec:3step}) 
provide viable alternative interpretations to the canonical model, where a scalar is resonantly produced through gluon fusion and decays
directly into two photons, $g g \rightarrow \varphi \rightarrow \gamma \gamma$.
In this section we discuss kinematic distributions that may differentiate between these three models.

The missing transverse energy ($\met$) may potentially distinguish the multi-step models, where there are final state particles that escape the detector, 
from the canonical model. In addition, the number of jets ($N_j$) observed in association with diphotons of invariant mass near the resonance may differentiate between 
the 3-step model, where the initial state is $s\bar s$, and the other models where the initial state is $gg$.
More generic information is provided by the transverse momentum distribution of the diphoton system ($p_{T\gamma\gamma}$), due to its sensitivity to anything in the event 
against which the two photons can recoil.

The $\met$, $N_j$ and $p_{T\gamma\gamma}$ distributions have been presented by the ATLAS Collaboration \cite{ATLAS-CONF-2016-018}, 
using 3.2 fb$^{-1}$ of data, for a diphoton invariant mass in the  window $m_{\gamma \gamma} \in [700,840] \GeV$. That choice was motivated by the large 
excess observed near  $m_{\gamma \gamma} = 750$ GeV \cite{ATLAS-CONF-2015-081,CMS:2015dxe,ATLAS-CONF-2016-018,CMS:2016owr}.
Even though that excess was not confirmed in later data \cite{ATLAS:2016eeo, Khachatryan:2016yec}, we compare the ATLAS kinematic distributions to 
the predictions of our multi-step models. This allows us  to estimate how small the mass splittings need to be in order for 
a cascade decay to be consistent with an initial interpretation of a diphoton resonance as being due to the canonical model.

\begin{figure}
\centering

\begin{subfigure}{0.7\textwidth}
\includegraphics[width=\textwidth]{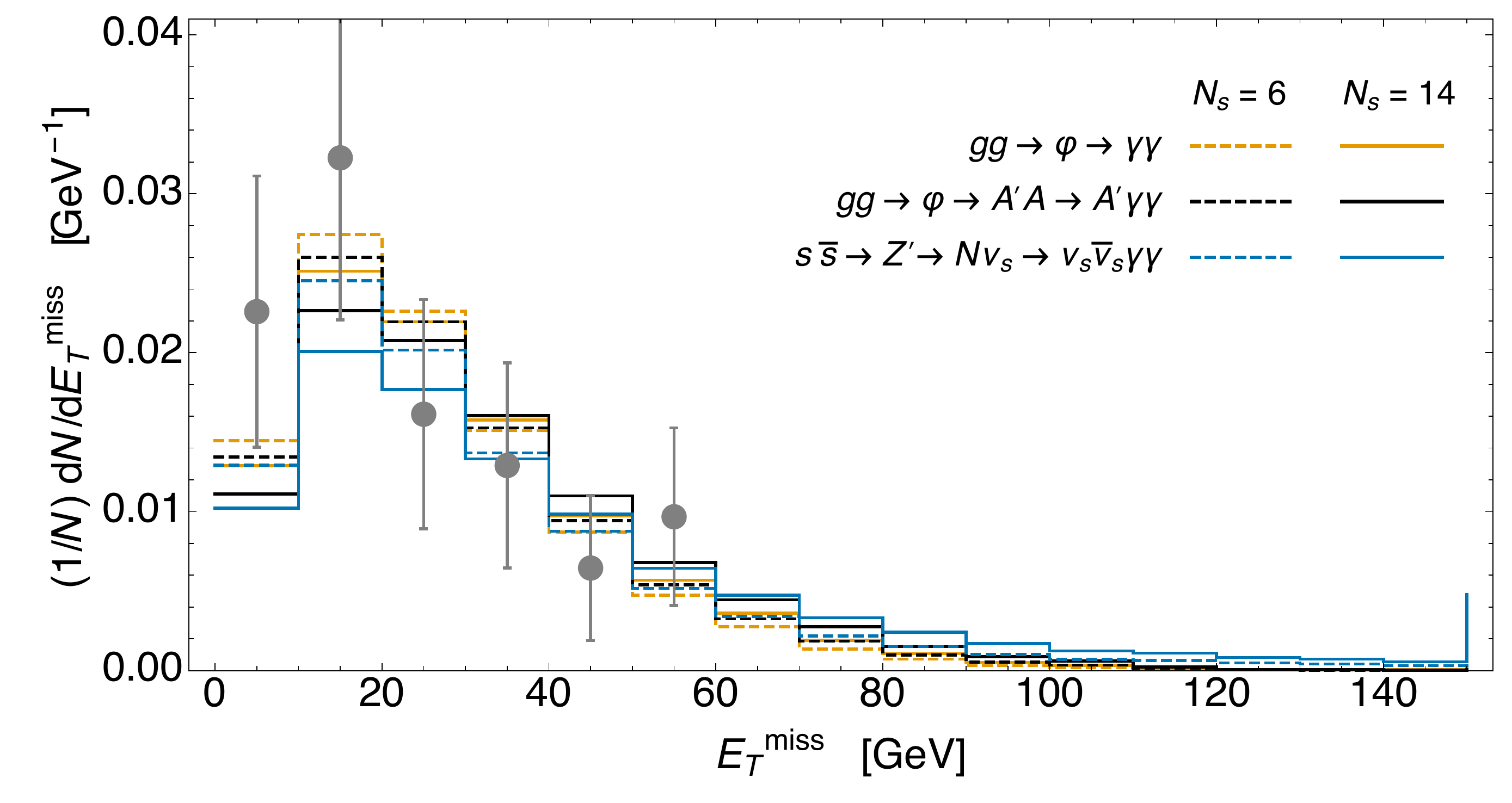}
\end{subfigure}

\begin{subfigure}{0.7\textwidth}
\centering
\includegraphics[width=\textwidth]{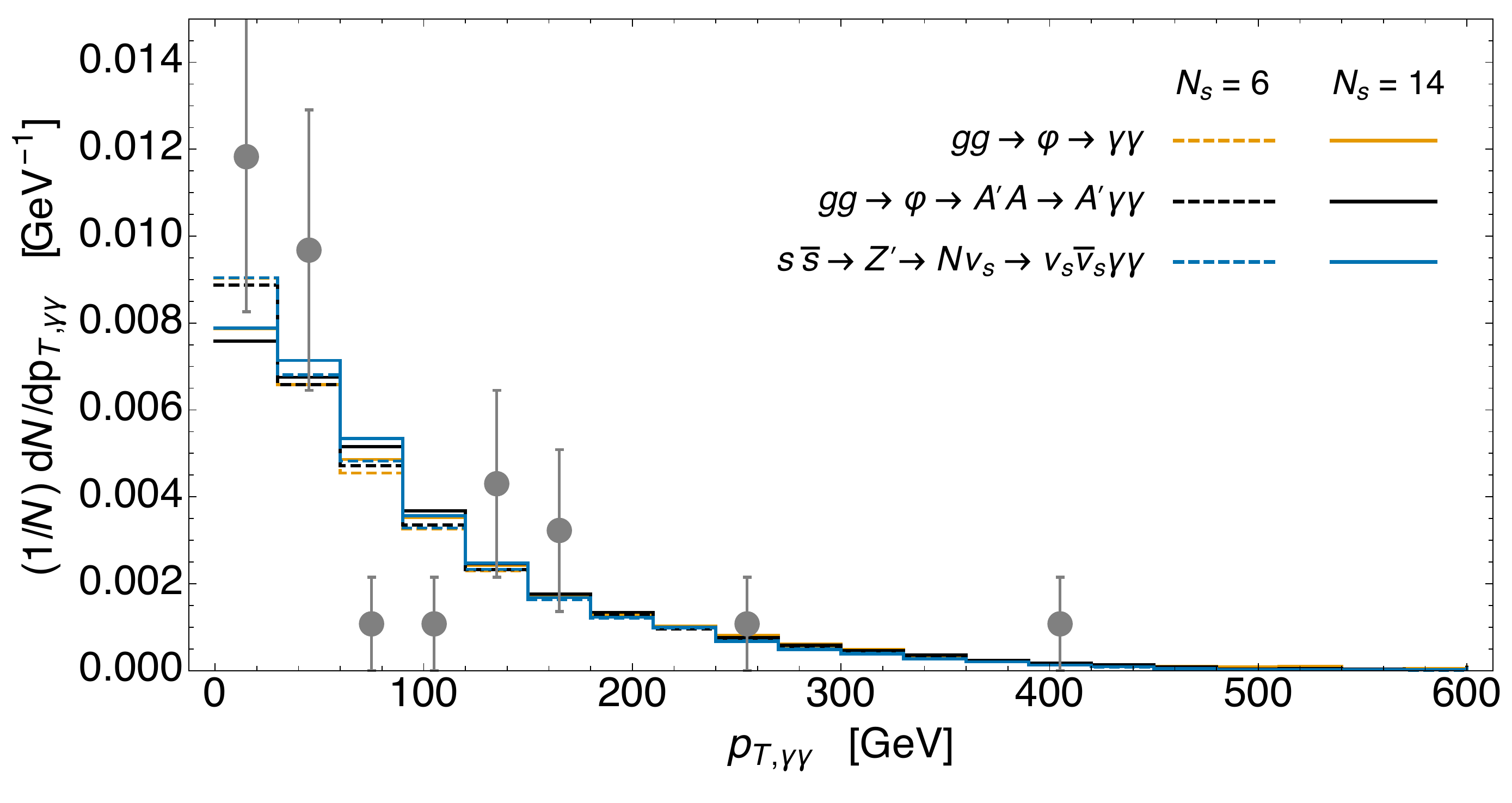}
\end{subfigure}

\begin{subfigure}{0.7\textwidth}
\centering
\includegraphics[width=\textwidth]{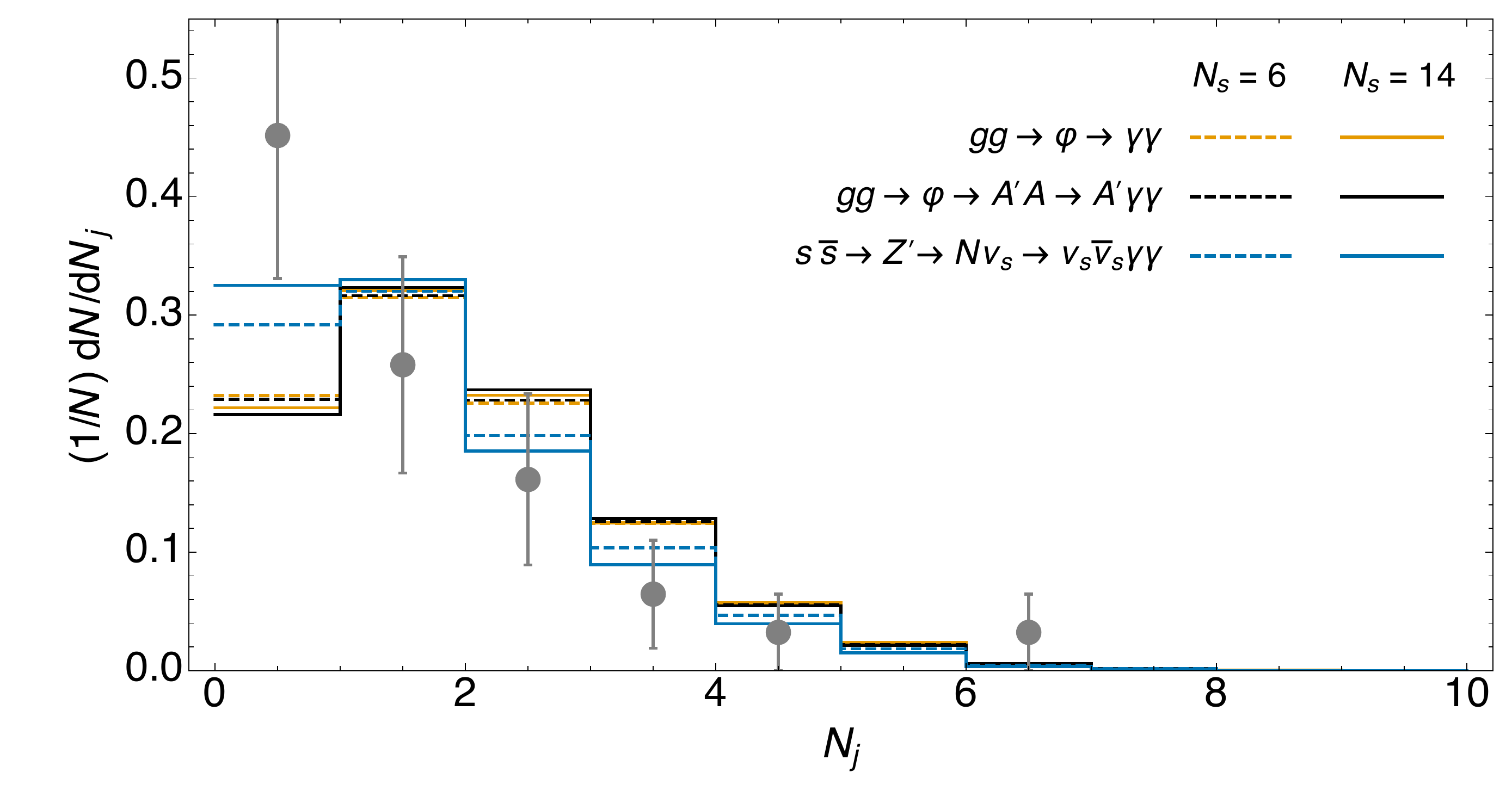}
\end{subfigure}

\caption{\label{fig:kindist} Normalized kinematic distributions for the 2-step, 3-step and canonical $g g \rightarrow \varphi \rightarrow \gamma \gamma$ models supposing 
17 background events and $6 \leq N_s \leq 14$ signal events.
The acceptance/cut efficiency is $\epsilon \simeq 0.6$, so that this choice corresponds to $3.1 \fb \leq \sigma_{\gamma \gamma} \leq 7.3 \fb$ for $3.2 \fb^{-1}$ of data.
Data (gray dots) are taken from \cite{ATLAS-CONF-2016-018}.
Top: Missing transverse energy; the uptick at the end corresponds to overflow, \ie, events with $\met > 150 \GeV$. Middle: $p_T$ of the $\gamma \gamma$ system. Bottom: Jet multiplicity; the overlap of the 2-step and canonical models (black and orange) is due to processes arising from the same initial state.}
\end{figure}

The three distributions are shown in \Fref{fig:kindist} and are generated as follows. Partonic events are generated in MadGraph \cite{Alwall:2014hca}, with showering carried out subsequently in \textsc{Pythia 6.4} \cite{Sjostrand:2006za}.  Detector simulation is carried out using \textsc{Delphes 3.3.0} \cite{deFavereau:2013fsa}.
For masses, we use the benchmark point 1 given in Eq.~(\ref{eq:bench}) 
for the 2-step model,
\be
M_{\varphi}  = 780 \GeV \;\; , \;\;  M_A  = 750 \GeV \;\; ,  \;\; M_{A'}  = 0 ~~,
\ee
and in Eq.~(\ref{eq:threestepexampleppoint})
for the 3-step model, 
\be
M_{Z'}  = 790 \GeV \;\; , \;\; m_{N}  = 760 \GeV  \;\; ,  \;\; M_A  = 750 \GeV \;\; ,  \;\; m_{\nu_s}  = 0  ~~.
\ee
For background, we take the distributions from \textsc{Sherpa} \cite{Gleisberg:2008ta} given by ATLAS \cite{ATLAS-CONF-2016-018}.
While the distributions shown here are for unmatched samples, the distributions are not significantly altered by matching.

Total expected distributions for each model are obtained by combining a weighted amount of the signal distribution to the background distribution. Based on the functional fit provided in \cite{ATLAS-CONF-2016-018}, which is in good agreement with recent theoretical calculations \cite{Campbell:2016yrh}, we take the expected number of background events in a putative signal window $m_{\gamma \gamma} \in [700,840] \GeV$ to be $N_b = 17$. We then add a signal corresponding to $N_s$ events and renormalize the distributions. In other words, we take
\be
\left(\frac{1}{N} \frac{d N}{d x}\right)_{\rm total} = \frac{N_b}{N_s + N_b} \left(\frac{1}{N} \frac{d N}{d x}\right)_{\rm background} + \frac{N_s}{N_s + N_b} \left(\frac{1}{N} \frac{d N}{d x}\right)_{\rm signal}
\ee
where $x$ represents a kinematic variable of interest.

Supposing the excess observed in \cite{ATLAS-CONF-2015-081,CMS:2015dxe,ATLAS-CONF-2016-018,CMS:2016owr} had been due to a state with mass 750 GeV decaying to $\gamma\gamma$, the 31
events observed by ATLAS in the signal window would have corresponded to $N_s \simeq 14$. While a potential signal had also been observed at CMS, the rate was somewhat lower. Thus, we consider a range of signal events $6 \leq N_s \leq 14$, and require events satisfy $700 \GeV \leq m_{\gamma \gamma} \leq 840 \GeV$ and pass the ATLAS cuts.
For simplicity, we take these to be $p_{T,\gamma_1} > 300 \GeV$ and $p_{T,\gamma_2} > 225 \GeV$, as for photons arising from a 750 GeV resonance,
but have confirmed that the given distributions are not particularly sensitive to this choice.
The acceptance and cut efficiency for the various models is $\epsilon \simeq 0.6$, implying that the above  $N_s$ range in  $3.2 \fb^{-1}$ of data 
corresponds to $3.1 \fb \leq \sigma_{\gamma \gamma} \leq 7.3 \fb$.
The lower limit of this range arises
from a fit to the combination of ATLAS and CMS datasets from both the 8 TeV run and the 2015 run at 13 TeV  \cite{Buckley:2016mbr}.

Note that the benchmark points given in Eqs.~(\ref{eq:bench}) and (\ref{eq:threestepexampleppoint}) have been chosen to yield smaller 
cross sections than the range discussed here, so that they are not ruled out by the large 2016 datasets \cite{ATLAS:2016eeo, Khachatryan:2016yec}.  
The larger $\sigma_{\gamma \gamma}$ assumed in this section corresponds to $\kappa \approx 1$ in the 2-step model and $g_z \approx 0.5$ in in the 3-step model;
these larger values have no impact on the shape of the kinematic distributions, which is the focus of this section.

As the two models described above feature additional particles that escape the detector, they exhibit somewhat longer tails in the $\met$ distribution compared to the canonical
 $gg \rightarrow \varphi \rightarrow \gamma\gamma$ model. 
Correspondingly, increasing $N_s$ shifts the distribution towards higher $\met$ for these models, as can be seen in \Fref{fig:kindist}.
This is especially true for the 3-step model as two $\nu_s$ escape the detector, leading to additional $\met$ and even an ``overflow'' of events with $\met > 150 \GeV$.
By comparison, the $\met$  distribution in the canonical model  does not change substantially.

Overall, our multi-step models with mass splittings below 30 GeV or so cannot be easily differentiated from the canonical model. 
Nevertheless, if a high-mass diphoton state were to be observed, then the kinematic distributions could immediately constrain the parameter spaces to exhibit such small splittings,
and with more data would discriminate between the various models.

In fact, depending on the cross section $\sigma_{\gamma \gamma}$, even smaller splittings may be required for the 3-step model to mimic the canonical model.
For $N_s = 6$ (14), this model predicts 3.1 (6.6) events with $\met > 60 \GeV$, whereas ATLAS observed 0. This tension with data would be alleviated by smaller mass splittings (and correspondingly more tuning), but there is also tension between a compressed spectrum and achieving sufficiently large branching fractions and rate. Moreover, as discussed in \Sref{sec:productionanddecay}, off-shell $Z'$ contributions will still yield events with non-negligible $\met$.
Thus, the 3-step model would more readily be disfavored by the non-observation of $\met$.
Though, we note that potential alternatives do exist for alleviating tension in the 3-step model, for instance if the $\nu_s$ state were to be somewhat massive and/or decay producing soft jets, perhaps in conjunction with another, lighter state that escaped the detector.
For comparison, the expected number of events with $\met > 60 \GeV$ is 1.3 (2.4) for the simple model and 1.7 (3.3) for the 2-step model, with background contributing 0.4 events.

Another difference between models appears in the $N_j$ distribution (see \Fref{fig:kindist}). Specifically, gluon-initiated processes (in the canonical and 2-step models) 
exhibit
higher jet multiplicities than the sea quark ($s \bar{s}$ and $b \bar{b}$)-initiated 3-step model. While both types of processes are consistent with the 
considered data, even this simple distribution may provide evidence for the production mechanism with a larger data set \cite{Ebert:2016idf}.

The $p_{T,\gamma\gamma}$ distributions, meanwhile, are less useful in distinguishing between these models, and do not change significantly with $N_s$ except at very low $p_{T,\gamma\gamma}$. This likely results from the small splittings required to satisfy the lack of observed $\met$, which produce smaller boosts for the state that ultimately decays to two photons.

\section{Conclusions}\label{sec:conclusions}

We have proposed two weakly-coupled models capable of giving rise to signals that could, at first glance, be interpreted as arising from a scalar produced in gluon fusion and decaying directly to two photons.
Both feature additional final state particles, resulting in missing transverse momentum that could ultimately be used to distinguish these models from this minimal interpretation.
These models provide viable examples of different initial states: gluons in the 2-step model (\Sref{sec:2step}) and sea quarks in the 3-step model (\Sref{sec:3step}). 
Differences in initial state radiation from quarks and gluons may appear in, \eg, jet multiplicity distributions.  Furthermore, the additional structure present in these models could be revealed by looking at other kinematic distributions such as $M_T$.  Although the limited kinematic information presented by ATLAS \cite{ATLAS-CONF-2016-018}, when interpreted as a possible observation, restricts the spectrum of our models to have small splittings, we have shown that the data are incapable of distinguishing the canonical 1-step from our 2-step and 3-step models.

A distinctive feature of our models is that the particle produced in the $s$-channel is different from the one that decays to $\gamma \gamma$.
This avoids the tension between simultaneously achieving a sufficiently large production via gluon fusion
and a sufficiently large diphoton branching fraction 
(overcoming the decay back to jets), 
which typically requires the particles running in the loops to have large charges or multiplicities 
(see, \eg, \cite{Franceschini:2015kwy}).

In the 2-step model, the particle produced in the $s$-channel ($\varphi$) and the particle that decays into photons ($A$) are the two components of a complex scalar field whose VEV is much smaller than its mass. Consequently, their masses are naturally almost degenerate, leading to little activity in the event besides the two high-$p_T$ photons.

In the 3-step model,  a spin-1 particle ($Z'$) is produced in the $s$-channel, and two spin-0 particles ($A$ and $\varphi$) decay into $\gamma\gamma$.
One notable feature of this model is that the process responsible for generating the signal does not involve loops of colored particles; production occurs at tree level, while the diphoton decay occurs via loops of ``anomalons'' required by the consistency of the theory.
Although all their masses are set by the VEV that breaks the gauged $U(1)$ symmetry, a mass splitting between the $Z'$ and the scalars of less than 10\% requires some tuning. 
While it could simply be a coincidence, there are some possible explanations for this small mass splitting.
For example, in models where multiple particles have masses proportional to a single scale of spontaneous symmetry breaking, mass relations could be the result of renormalization 
group fixed ratios as in, \eg, \cite{Kearney:2013xwa}. A small mass splitting between $A$ and $\varphi$ is natural, and could potentially lead to the two resonances being initially observed as a single, wide resonance.

We also highlight that, depending on the spins of the particles involved, off-shell contributions to the diphoton rate can be significant in certain models because of relatively small splittings. It is important to take these effects into account as they can significantly alter kinematic distributions relative to the case where all intermediate state particles are on-shell.

There exist other exotic possibilities that may initially be misinterpreted as a directly-produced diphoton resonance.  For instance, a resonantly-produced state could decay to two pairs of highly-collimated photons resulting from the decay of a light intermediate state \cite{Dobrescu:2000jt}.  Beyond the diphoton final state, this work seeks to emphasise that any resonance-like signal requires careful analysis from all angles to confirm the true nature of the underlying model, and in particular that a multi-step process can to some extent mimic a single $s$-channel resonance.

\bigskip

\bigskip

\bigskip

{\it \bf Acknowledgments:} 
BD and PJF would like to thank University of Oregon, and the organizers of the workshop ``Emerging New Physics at the LHC,'' for hospitality during the very final stages of this work. 
This work was supported by the DoE under contract number DE-SC0007859 and Fermilab, operated by Fermi Research Alliance, LLC under contract number DE-AC02-07CH11359 with the United States Department of Energy.

\vfil
\end{document}